\documentclass[11pt]{article}\topmargin=-0.5cm\hfuzz=10pt  \oddsidemargin=-0.1cm \textheight 222mm\textwidth=16.3cm
\usepackage{amssymb}
\newcommand{\comm}[1]{}

\usepackage{pdfpages}

\def\short{\comm}

\usepackage{color}

\usepackage{endnotes}

\newtheorem{theorem}{Theorem}
\newtheorem{lemma}{Lemma}
\newtheorem{corollary}{Corollary}
\newtheorem{definition}{Definition}

\def\e{\varepsilon}

\def\defi{\stackrel{{\scriptscriptstyle \Delta}}{=}}
\def\defi{:=}

\def\a{\alpha}
\def\d{\delta}
\def\o{\omega}
\def\O{\Omega}

\def\Y{{\cal Y}}

\def\w{\widehat}
\def \Ind{{\,\rm Ind\,}}
\def \Ind{{\mathbb{I}}}

\def\Re{{\rm Re\,}}
\def\Im{{\rm Im\,}}

\def\R{{\bf R}}
\def\E{{\bf E}}

\def\Z{{\cal Z}}
\def\PP{{\cal P}}

\def\b{\beta}

\def\C{{\bf C}}

\def\ww{\widetilde}
\def\X{{\cal X}}
\def\t{\theta}
\def\oo{\bar}

\def\p{\partial}

\newcommand{\be}{\begin{equation}}
\newcommand{\ee}{\end{equation}}
\newcommand{\bd}{\begin{displaymath}}
\newcommand{\ed}{\end{displaymath}}
\newcommand{\ba}{\begin{array}{ll}}
\newcommand{\ea}{\end{array}}
\newcommand{\baa}{\begin{eqnarray}}
\newcommand{\eaa}{\end{eqnarray}}
\newcommand{\baaa}{\begin{eqnarray*}}
\newcommand{\eaaa}{\end{eqnarray*}}

\def\PP{{\cal P}}

\def\oo{\bar}

\def\a{\alpha}

\def\ew{\left(e^{i\o}\right)}

\def\ZZ{{\mathbb{Z}}}

\def\TT{{\mathbb{T}}}
\def\zz{{\bf h}_1}
\def\zzm{{\bf h}_{1,m}}
\def\XHF{\X(\O)}
\def\XHFo{\X(\oo\O)}
\def\g{\gamma}
\def\G{\Gamma}
\def\XHFF{\X(\O_0)}

\def\ee{\e}

\date{ Submitted:  25.02.2023 }
\date{Submitted:  21.02.2023;  revised:  28.03. 2023}
\title{Near-ideal predictors and causal filters for discrete time signals \comm{based on causal  approximations  } }
\author{Nikolai Dokuchaev}
 
\begin{document}
 \def\short{\comm}
\def\brea{}
\def\breakk{}
\def\break{}
\def\BRR{}
\def\breakm{\nonumber\\  }\def\BR{}\def\BRR{}
\def\breacm{}
\def\dt{}
\maketitle
\let\thefootnote\relax\footnote{{Accepted to {\em Problems of Information Transmission}}}

\%\begin{abstract} \end{abstract}


\begin{abstract} The paper presents   linear  predictors and causal filters for discrete  time  signals  featuring some different kinds of  spectrum degeneracy. These predictors and filters  are based on approximation of ideal non-causal transfer functions   by causal transfer functions represented  by polynomials of Z-transform of the unit step signal.  
\end{abstract} 

{\bf Key words}:  discrete time signals,  forecasting, predictors,  filters, causal transfer functions,  causal approximation,  high frequency signals, low frequency signals.

\section{Introduction}
It is well known that certain
degeneracy  on the  spectrum can ensure 
opportunities for prediction and interpolation of the signals;
see, e.g.,  \cite{B}-\cite{V}. The present paper considers  discrete time signals in the deterministic setting, where  only a single 
trajectory  of the signal is observed,  rather than a set of samples of trajectories that would allow to apply statistical methods.  The method that we use is based  on the frequency analysis. It is known in principle that these signals  are predictable, i.e., they allow  unique extrapolations from their past observations, if they have a finite  spectrum gap, i.e. a segment  of the unit circle 
$\TT=\{z\in\C: \ |z|=1\}$, where their  Z-transform vanish; see, e.g.  \cite{D12}.
This gap can be arbitrarily small and can be even reduced to a point, under certain 
conditions on the rate of spectrum degeneracy in a neighbourhood of this point. Respectively, an ideal  low-pass filter or high-pass-filter would convert  a non-predictable signal to predictable one. This is why  
these ideal filters cannot be causal. 

For discrete time signals, some predictors    based on irrational  causal transfer functions were obtained  in \cite{D12,D12b}.  The corresponding transfer functions were presented  via exponentials of rational  functions or power functions. 
In  \cite{D16},  some low-pass filters were also constructed based on a similar principle.

The paper addresses again the prediction and filtering problems for discrete time signals; it offers  new  predictors and causal filters approximating ideal filters. The causal transfer functions for these predictors and filters are represented as polynomials of  Z-transform of the unit step signal, i.e., polynomials of  $(1-z^{-1})^{-1}$. For  the predictors, the corresponding transfer functions  approximate the function $e^{i\o T}$ on $\TT$, where $\o\in (-\pi,\pi]$ represents the frequency, and where  an integer $T>0$ represents a  preselected prediction horizon. For the filters, the corresponding transfer functions  approximate the  real valued step function representing the trace on $\TT$ of Z-transform of an ideal filter. The approximation is possible for signals  with some arbitrarily small spectrum gap; the resulting signal could have   a wider preselected spectrum gap. 
This polynomial approximation  method is based  on the approach developed in \cite{D22,D22rus}  for prediction of  continuous time signals.

 The results  are   applicable  for high frequency signals as well as 
 for signal for a spectrum gap located anywhere on $\TT$, for example,  low frequency signals.  Moreover, the paper shows that  some signals with a non-degenerate spectrum also can be predicted in a half of the timeline given some conditions on some spectrum type characteristics of the trace on this half of timeline.

  These new predictors and filters allow an explicit representation in the time domain and  in the frequency domain; in addition,  they  are independent on the
spectral characteristics of the input signals with given type of the spectrum  degenerocity. 
Some computational approach based on model fitting is suggested.

The paper is organized in the following manner. In Section
\ref{secDef}, we formulate the definitions. In Section
\ref{secM}, we formulate
 the main theorems on predictability and predictors (Theorem \ref{ThP} and Theorem \ref{ThF}). In Section \ref{secR}, we discuss  representation of transfer functions in the time domain. In Section \ref{secD}, we discuss  some implementation problems.   
    In Section \ref{secExp}, a method of computing  approximating functions for exponentials $e^{i\o}$ is suggested.
In Section \ref{secLF}, we suggest  extension of the results on low frequency and other 
  signals.
 Section \ref{secProof} contains the proofs.

\section{Problem setting }\label{secDef}

\subsection*{Some notations}

Let $\ZZ$ be the set of all integers.

We denote by $\ell_r$ the set of all functions (signals) $x:\ZZ\to \C$, such that
$\|x\|_{\ell_r}\defi \left(\sum_{t=-\infty}^{\infty}|x(t)|^r\right)^{1/r}<+\infty$
for $r\in[1,\infty)$.

\par
For  $x\in \ell_1$ or $x\in \ell_2$, we denote by $X=\Z x$ the
Z-transform  \baaa X(z)=\sum_{t=-\infty}^{\infty}x(t)z^{-t},\quad
z\in\TT. \eaaa Respectively, the inverse Z-transform  $x=\Z^{-1}X$ is
defined as \baaa x(t)=\frac{1}{2\pi}\int_{-\pi}^\pi
X\left(e^{i\o}\right) e^{i\o t}d\o, \quad t=0,\pm 1,\pm 2,....\eaaa
If $x\in \ell_2$, then $X|_\TT$ is defined as an element of
$L_2(\TT;\C)$.

We denote by $\Ind$  the indicator function.
\subsection*{Some definitions}
 
Let either  $E=\R$ or $E=\C$.

Let $ \X\subset \ell_\infty$ be a set currently observable discrete time signals 
 with values in $E$. 

Let $\PP(\X)$ be the set of all continuous mappings $p:\X\to \ell_\infty$ such that, for any $x_1,x_2\in\X$ and $\oo t\in\ZZ$, we have that 
 $p(x_1(\cdot))(t)=p(x_2(\cdot))(t)$ for all $t\le\oo t$   if $x_1(t)=x_2(t)$ for all $t\le \oo t$. In other words, this is the set of "causal" mappings;
we will look for predictors and filters in  this class.

Let us consider first a prediction problem. Let  an integer $T\ge 1$ be given.
The goal is to estimate, at current times $t$, the values
$x(t+T)$, using historical values of the observable process
$x(s)|_{s\le t}$. Therefore,   $T$ is the  prediction horizon in this setting.

\begin{definition}\label{defP} Let $\X\subset\ell_\infty$ and $\tau\in\{-1,0,+\infty\}$.
\begin{itemize} 
\item[(i)]
 We say that the  class $\X$ is predictable with the prediction horizon $T$ up to time $\tau$  if there exists a sequence $\{\ww
p_{d}(\cdot)\}_{d=1}^{+\infty}\subset\PP(\X)$ such that $$
\sup_{t\in\ZZ,\ t\le \tau -T}|x(t+T)-\ww y_{d}(t)|\to 0\quad \hbox{as}\quad
d\to+\infty\quad\forall x\in\X, $$ where \baaa &&
\ww y_{d}= \ww p_d(x(\cdot)).\label{predict} \eaaa 
\item[(ii)] 
 We say that the class $\X$ is  uniformly  predictable 
  with the prediction horizon $T$ up to time $\tau$  if   there exists a sequence $\{\ww
p_{d}(\cdot)\}_{d=1}^{+\infty}\subset\PP$  such that 
\baaa \sup_{t\in\ZZ, \ t\le \tau-T}|x(t+T)-\ww
y_{d}(t)|\to 0\quad \hbox{as}\quad
d\to+\infty\quad\hbox{uniformly in} \quad x\in\X,\eaaa 
where  $\ww
y_{d}(\cdot)$ is as in part (i) above.
\end{itemize}
\end{definition}
Functions $\ww y_{d}(t)$ in the definition above
can be considered as approximate predictions of the process $x(t+T)$.

Let us consider now the filtering problem. 

Let $\O\in  (0,\pi)$   be given. Let  a function $\Phi_\O:\TT\to\R$ be defined such that $\Phi_\O\ew=\Ind_{|\o|\ge \O}$.

We consider an ideal high-pass filter such that the trace of its  transfer function on $\TT$ is $\Phi_\O\ew$, $\o\in(-\pi,\pi]$, i.e., filters with the  the suppression interval    $(-\O,\O)$.

The goal is to find  an arbitrarily close approximation of  this  non-causal transfer function  $\Phi_\O\ew$ by causal transfer functions.

\begin{definition}\label{defF} Let $\X\subset\ell_\infty$.
\begin{itemize}  
\item[(i)] 
 We say that a class $\X\subset \ell_2$  allows causal high-pass filtering 
 with the suppression interval
  $(-\O,\O)$  if there is  a sequence $\{\ww
p_{d}(\cdot)\}_{d=1}^{+\infty}\subset\PP(\oo\X)$ such that $$
\sup_{t\in\ZZ}|x(t+T)-\ww y_{d}(t)|\to 0\quad \hbox{as}\quad
d\to+\infty\quad\forall x\in\X, $$ where \baaa &&
\ww x=\Z^{-1}(\Phi_\O X),\quad  \ww y_{d}= \ww p_d(x(\cdot)).\label{predictF} \eaaa  
\item[(ii)] 
 We say that the class $\X$ allows uniform  causal high-pass filtering  with   the suppression interval 
  $(-\O,\O)$  if   there exists a sequence $\{\ww
p_{d}(\cdot)\}_{d=1}^{+\infty}\subset\PP$  such that \baaa \sup_{t\in\ZZ}|\ww x(t)-\ww
y_{d}(t)|\to 0\quad \hbox{as}\quad
d\to+\infty\quad\hbox{uniformly in} \quad x\in\X,\eaaa 
where $\ww x$ and  $\ww
y_{d}$ are as in part (i) above.
\end{itemize}
\end{definition}
In the last definition,  operators $p_d$ represent causal near-ideal high pass filters; they ensure,
for the class $\X$, an arbitrarily close approximation of  the non-causal ideal high-pass filter defined by its transfer function $\Phi_\O$.

\section{The main result} 
\label{secM}

\comm{Let $\UHF$ be the set of 
signals $x\in \XHF$ such that  $\int_{-\pi}^\pi|X\ew |d\o \le 1$
for  $X=\Z x$.}

For $d=0,1,2,...$, let $\Psi_d^E$    be the set of all  functions $\psi:\C\setminus \{1\}\to\C$ represented as  \baa
\psi(z)=\sum_{k=0}^d \frac{a_k}{(1-z^{-1})^k},
\label{g}\eaa  where $a_k\in E$ can be any.  Let $\Psi^E\defi \cup_{d}\Psi_d^E$.

\vspace{0.4cm}   

\begin{lemma}\label{lemmaA}  For $\oo\O\in (0,\pi)$, let the function $\zeta:[-\pi,\pi]\to \C$ be defined either as $\zeta(\o)=e^{i\o T}$  or as $\zeta(\o)=\Ind_{|\o|\ge \oo\O}$. Then,
for any  $\ee>0$,  there exists  a integer $d=d(\ee,T)>0$ and $\psi_d\in\Psi_d^\R$ such that 
\baa
\sup_{\o\in [-\pi,\pi]:\ |\o|\ge \oo\O} 
|\zeta(\o)-\psi_d\ew|  \le \ee. \label{e2}
\eaa
\end{lemma}

For $\oo\O\in  (0,\pi)$,  let 
 $\XHFo$   be the set of all signals $x:\ZZ\to E$ such that 
$x(\cdot)\in\ell_2$ and $X\ew=0$ for $\o\in (-\oo\O,\oo\O)$ and  $X=\Z x$.

Further,  for $\tau=-1,0$,  let $\X(\tau,\oo\O)$ be the set of all real signals $x:\ZZ\to\R$ such that 
$x(\cdot)\in\ell_2$ and the following holds:

\begin{itemize}
\item If $\tau=0$, then \baa
2\sum_{t=-\infty}^{-1} \cos(\o t)x(t)+x(0) =0\quad \hbox{for} \quad \o\in (-\oo\O,\oo\O).
\label{L1}\eaa
\item If $\tau=-1$, then \baa
\sum_{t=-\infty}^{-1} \sin(\o t)x(t)=0\quad \hbox{for} \quad \o\in (-\oo\O,\oo\O).\label{L2}
\eaa
\end{itemize}

We say that the processes from $\X(\tau,\oo\O)$  described above feature will 
 call  a left-sided spectrum degeneracy. 

The  feature of the processes from $\X(\tau,\oo\O)$, $\tau=-1,0$,  described above, we  will 
 call  a left-sided spectrum degeneracy. 
In addition, we define $\X(\infty,\oo\O)$ as $\X(\oo\O)$.

\begin{theorem}\label{ThP} Let $\O\in (0,\pi)$ be given, $\tau\in\{-1,0,+\infty\}$.
 The predictability up to time $\tau$ for   $x\in \X(\tau,\O)$ 
considered in Definition \ref{defP}(i), as well as the  uniform predictability  up to time $\tau$ for   $x\in  \X(\tau,\O)\cap \{x\in \ell_2:\ \|x\|_{\ell_2}\le 1\}$  
considered in Definition \ref{defP}(ii),
 can be ensured with
 the sequence of the predictors $ p_d:\XHF\to \ell_2$, $d=1,2,....,$ defined by their transfer functions $\psi_d (z)$ selected as in Lemma \ref{lemmaA} with $\zeta(\o)=e^{i\o T}$. More precisely, for any $\oo\e>0$ and  $\w y_d(t)= p_d(x(\cdot))(t)$, the estimate  
\baaa
\sup_{t\in\ZZ,\ t\le \tau-T}|x(t+T)-\w y_d(t)|\le\e
\eaaa
holds if   $d$ and $\psi_d$ are such that
(\ref{e2}) holds with $\zeta(\o)=e^{i\o T}$ for sufficiently small  $\e$. 
\end{theorem}
\begin{theorem}\label{ThF} For  $\O\in (0,\pi)$ and any $\O_0\in (0,\O)$,  for   $x\in \X(\O_0)$,
the causal filtering considered in Definition \ref{defF}(i), as well as the uniform causal filtering  for   $x\in \X(\O_0) \cap \{x\in \ell_2:\ \|x\|_{\ell_2}\le 1\}$ considered in Definition\ref{defF}(ii)
 can be ensured with
 the sequence of the causal filters $ p_d:\XHF\to \ell_2$, $d=1,2,....,$ defined by their transfer functions $\psi_d (z)$  selected as in Lemma \ref{lemmaA} with $\zeta(\o)=\Ind_{|\o|\ge \O}$. More precisely, for any $\oo\e>0$ and  $\w y_d(t)= p_d(x(\cdot))(t)$ and $\ww x=\Z^{-1}(\Phi_\O X)$, the estimate  
\baaa
\sup_{t\in\ZZ}|\ww x(t)-\w y_d(t)|\le\e\eaaa
 if   $d$ and $\psi_d$ are such that
(\ref{e2}) holds with $\zeta(\o)=\Ind_{|\o|\ge \O}$ for sufficiently small  $\ee$. 
\end{theorem}
According to this theorem,  a process with an arbitrarily small spectrum gap $(-\O_0,\O_0)$ 
can be converted, using causal operations,  into a process with larger spectrum gap  up to $(-\O,\O)$.

\par 

It van be noted that:
\begin{itemize} 
\item
The transfer functions $\psi_d(z)$ are analytic in the domain
$\C\setminus\{1\}$.  
If we apply their traces  $\psi_d \ew|_{\o \in (-\pi,\pi]}$ on $\TT$ for calculation of the outputs 
for inputs $x\in \XHF$, then we obtain the same outputs as for the  functions $\psi_d \ew \Ind_{\o\in (-\pi,\pi], |\o|\ge\O}$.  
\item
For real valued inputs $x$, the outputs of these predictors and filters are real valued.
\item  $p_d(\cdot)$ depends on  $T$ and $\O$  via the coefficients $a_k$ 
in the setting of  Theorem \ref{ThP}, and $p_d(\cdot)$ depends on  $\O$ and $\O_0$  via the coefficients $a_k$ 
in the setting of  Theorem \ref{ThF}. 
\end{itemize}
\section{Representation of operators $p_d(\cdot)$  in the time domain}
\label{secR}
Let either  $\oo\O=\O$ or $\oo\O=\O_0$.

Consider operators $h_k$ defined on $\XHFo$  by their transfer functions 
$H_k(z)=(1-z^{-1})^{-k}$, $k=0,1,2,...$\,. In other words, if $y=h_k(x)$ for $x\in\XHFo$, then 
$Y(z)=(1-z^{-1})^{-k}X(z)$ for  $Y=\Z y$ and $X=\Z x$.
 Clearly,  \baaa
&& H_{k+1}(z)=H_1(z)H_{k}(z),  \quad  h_{k+1}(x(\cdot))=h_1(h_{k}(x(\cdot))),  \quad k=0,1,2,3,...
\label{hint}\eaaa
Hence  $h_{k}(x(\cdot))\in\XHFo$ for all $k=0,1,2,...$, $x\in\XHFo$.
Therefore, 
 the 
Z-transforms of processes
$h_{k}(x(\cdot))$ vanish on $\{e^{i\o},\  \o\in[-\pi,\pi],\ |\o|<\oo\O\}$,   and the operators 
$h_k:\XHFo\to \ell_2$ are continuous, assuming that $\XHFo$ is a subspace of $\ell_2$ provided with $\ell_2$-norm.

Let  $\zz\in\ell_\infty$ be defined such that $\zz(t)=0$ for $t<0$ and  $\zz(t)=1$ for $t\ge 0$,
i.e. $\zz=\Z^{-1}H_1(z)$.

Let $I_{\oo\O}\defi [-\pi,-\oo\O]\cup  [\oo\O,\pi] $ and  $x\in\XHFo$.

Let us show that, in the time domain, the operator $h_1(x(\cdot))$ can be represented via causal convolution with the kernel $\zz$, i.e.
 if $x\in\XHFo$ then $h_1(x(\cdot))(t)=\sum_{s=-\infty}^t x(s)$.

Let $\zzm(t)=\zz(t)\Ind_{\{t<m-1\}}$. Clearly,  $\zzm\in\ell_2$. Let 
\baaa
H_{1,m}(z)\defi \Z \zzm=\frac{1-z^{-m}}{1-z^{-1}}, \quad R_{m}(z)\defi \Z (\zz-\zzm)=\frac{z^{-m}}{1-z^{-1}}.
\eaaa  Clearly, $(1-e^{-i\cdot})^{-1}e^{i \cdot} X(e^{i\cdot}) \in L_2(I_{\oo\O},\C)$ for any $t$. Hence
\baaa
\int_{-\pi}^\pi R_{m}\ew e^{i \o t}X\ew d\o= \int_{I_{\oo\O}} \frac{e^{-i m \o}}{1-e^{-i\o}} e^{i \o t}X\ew d\o \to 0
\quad \hbox{as}\quad m \to+\infty
\eaaa
for each  $t\in\ZZ$.  It follows  that if $x\in\XHFo$ then \baaa
h_1(x(\cdot))(t)=\sum_{s=-\infty}^t x(s) =\lim_{m\to +\infty}\Z^{-1} (R_m+\Z\zzm)(t)=\lim_{m\to +\infty}\sum_{s=-m}^t x(s),
\eaaa
and the series  converges for each $t\in\ZZ$.

This implies that
\baaa
h_{k}(x(\cdot))(t)=\sum_{s=-\infty}^t (h_{k-1}(x(\cdot))(s), \quad k=1,2,3,...\ .
\label{hint2}\eaaa

Therefore, the operators $p_d$ in Theorems \ref{ThP}-\ref{ThF}  can be represented as  \baaa
  p_d(x(\cdot))(t)=\sum_{k=0}^d a_k h_k(x(\cdot))(t),
  \label{Kkk}
\label{pred2} 
\eaaa
where 
\baa
\hspace{-0.5cm}&&h_k(x(\cdot))(t)\breakk= \sum_{s_{k-1}=-\infty}^{t}\
\sum_{s_{k-2}=-\infty}^{s_{k-1}}\ ... \sum_{s_1=-\infty}^{s_2}
\sum_{s=-\infty}^{s_1} x(s).
\label{int1}\eaa
All  series here converge as described above for $h_1$.

It can be noted that  $x\in\XHFo\cap \ell_1$ then the series  $\sum^t_{s=-\infty} x(s)$ converges absolutely; however,  for general  type   $x\in\XHFo$, there is no guarantee that  $x\in \ell_1$
or $ h_k(x(\cdot)) \in \ell_1$. 
 \section{ On numerical implementation of Theorems \ref{ThP}-\ref{ThF} }\label{secD}
The direct implementation of the predictors introduced in  Theorems \ref{ThP}-\ref{ThF}  requires 
evaluation of sums for   semi-infinite series that is not practically feasible.  
However, these theorems could lead to predicting methods  bypassing  this calculation.
Let us discuss these possibilities.
\par
 
Let  $t_1\in\R$ be given such that $t_1<\tau$, where $\tau$ in the setting of Theorem \ref{ThP} is such as described therein, and  $\tau=+\infty$ in the setting of Theorem \ref{ThF}. Let $x_k\defi h_k(x)$ for  $x\in\XHFF$,  $k=1,2,...$,  and let 
\baaa
\eta_k\defi x_k(t_1-1).
\eaaa
\begin{lemma}\label{lemma1}  In the notation of Theorems \ref{ThP}-\ref{ThF}, for any $t$ such that $t_1\le t<\tau+1$, we have  that   $\w y_d=p_d(x(\cdot))$ can be represented as
\baa
\w y_d(t)= a_0 x(t)+
\sum_{k=1}^d a_k \left(\sum_{l=1}^k c_{l}(t)\eta_l+f_{k}(t)\right).
\label{viaeta0}\eaa
Here $a_k\in\R$  are the coefficients for $\psi_d(z)=\sum_{k=1}^d a_k (1-z^{-1})^{-k}$   from Theorems \ref{ThP}-\ref{ThF}, 
\baaa
&& 
f_k(t)=\sum_{\tau_1=t_1}^{t_{}} \sum_{\tau_2=t_1}^{\tau_1} ...\sum_{s=t_1}^{\tau_k} x_{0}(s),
\eaaa
\baaa
&& c_{1}(t)=\sum_{\tau_1=t_1}^{t_{}} \sum_{\tau_2=t_1}^{\tau_1} ...\sum_{\tau_{k}=t_1}^{\tau_{k-1}}(\tau_{k}-t_1+1),\qquad c_{2}(t)=\sum_{\tau_1=t_1}^{t_{}} \sum_{\tau_2=t_1}^{\tau_1} ...\sum_{\tau_{k-1}=t_1}^{\tau_{k-2}}(\tau_{k-1}-t_1+1),\eaaa
and
\baaa
 &&c_{l}(t)=\sum_{\tau_1=t_1}^{t_{}} \sum_{\tau_2=t_1}^{\tau_1} ...\sum_{\tau_{l-1}=t_1}^{\tau_{k-l}}(\tau_{l-1}-t_1+1),\qquad l= 1,2,...,k-2,
 \eaaa
\baaa
&& c_{k-1}(t)=\sum_{\tau_1=t_1}^{t_{}}(\tau_{1}-t_1+1),\qquad  c_{k}(t)=t-t_1+1.
\eaaa
\end{lemma}

This lemma shows that calculation of $\w y_d(t)=p_d(x(\cdot))(t)$  is easy for $t>t_1$  if we know all $\eta_k$, $k=1,...,d$, and observe $x(s)|_{s=t_1,...,t}$.

Let us discuss some ways  to evaluate $\eta_k$ bypassing summation of infinite series. 

First, let us observe that (\ref{viaeta0})  implies a useful property given below.
\begin{corollary}\label{corr1} For any $\e>0$, there exist
 an integer $d=d(\e)>0$ and $a_0,a_1,....,a_d\in\R$  such that,
for any  $t_1\in \ZZ$,
there exist  $\oo\eta_1,\oo\eta_1,...,\oo\eta_d\in\R$  such that 
 $|\ww x(t)-y_d(t)|\le \e$  for all $t\ge t_1$, where \baa
 y_d(t)= y_d(t,x(t),\oo\eta_1,...,\oo\eta_d)\defi
 a_0 x(t)+\sum_{k=1}^d a_k \left(\sum_{l=1}^k c_{l}(t)\oo\eta_l+f_{k}(t)\right).
\label{viaeta}\eaa
\end{corollary}
\vspace{1mm}
In this corollary, $d=d(\e)$ and $a_k\in\R$ are such   as defined  in Theorems \ref{ThP}-\ref{ThF}.
\par
\subsubsection*{The case of prediction problem: Theorem \ref{ThP} setting} 
Let us discuss using  (\ref{viaeta0}) and (\ref{viaeta})  for evaluation of $\eta_k$ in Theorem \ref{ThP} setting. 

Let $\t>t_1$ and $\t<\tau+1$. Assume first  that  the goal is to forecast the value $\ww x(t)=x(t+T)$ 
 given observations at times $t\le  \t$,  in the setting of Theorems \ref{ThP}.  
It appears that if  $\t>t_1+T$ then Corollary \ref{corr1} gives an opportunity to construct predictors  via fitting  parameters  $\eta_0,...,\eta_d$ using past observations available for $t=t_1,...,\t-T$: we  can match  the values  $y_d(t,x(t),\oo\eta_1,...,\oo\eta_d)$ with the
 past observations $x(t+T)$. Starting from now, we assume that $\t>t_1+T$.

Let $d$ be large enough  such that
 $x(t+T)$ is approximated  by $\w y_d(t)$ as described in Theorem \ref{ThP}, i.e.,
 $\sup_{t\in\ZZ}|x(t+T)-\w y_d(t)|\le \e$  for some sufficiently small 
   $\e>0$, for some  choice of  $a_k$.

 As an approximation of the true $\eta_1,...,\eta_d$, 
we can accept a set $\oo\eta_1,...,\oo\eta_d$ such that
 \baa
|x(t+T)-y_d(t,x(t),\oo\eta_1,...,\oo\eta_d)|\le \e,\quad  t=t_1,...,\t-T.
 \label{eta1}\eaa
 (Remind that, at time $\t$,  values  $x(t+T)$ and $y_d(t,x(t),\oo\eta_1,...,\oo\eta_d)$
 are observable  for these $t=t_1,....,\t-T$).  If (\ref{eta1}) holds, we can conclude that
  $y_d(t,x(t),\oo\eta_1,...,\oo\eta_d)$ delivers an acceptable prediction of $x(t+T)$ for these $t$.
 Clearly, Theorem \ref{ThP} implies that a set $\oo\eta_1,...,\oo\eta_d$ ensuring (\ref{eta1})  
  exists since this inequality holds with $\oo\eta_k=\eta_k$.
 
 The corresponding value  $y_d(\t,x(t),\oo\eta_1,...,\oo\eta_d)$ would give an estimate for $\w y_d(\t)$ and, respectively, for $x(\t+T)$.

Furthermore,  finding a set $\oo\eta_1,...,\oo\eta_d$ that ensures (\ref{eta1})  could still be difficult.  Instead, one can consider fitting predictions and observations at a finite  number of points $t=t_1,...,T-\t$.

Let  a integer $\oo d\ge d$ and  a set $\{t_m\}_{m=1}^{\oo d}\subset \ZZ$ be selected such that $t_1<t_2<t_3<...<t_{\oo d-1}<t_{\oo d}\le\t-T$.  We suggest to  use observations  $x(t)$ at times $t=t_m$. 
Consider  a system of equations
\baa
a_0 x(t_m)+\sum_{k=1}^d a_k \left( \sum_{l=1}^k c_{l}(t_m)\oo\eta_l+f_{k}(t_m) \right)=\zeta_m, \quad m=1,...,\oo d.\label{sys3} \eaa
\par
Consider first the case where $\oo d=d$. In this case, we can select $\zeta_m= x(t_m+T)$; these values are directly observable, without calculation of semi-infinite series required for $\w y_d(t_m)$.  The corresponding choice of $\oo\eta_k$ ensures zero prediction error for $x(t_m+T)$, $m=1,...,\oo d$. 

Including into consideration more observations, i.e.,  selecting larger $\oo d>d$ and larger set $\{t_1,....,\t-T\}$, would
 improve estimation of $\eta_k$.
  If we consider $\oo d>d$, then, in the general case,  it would not be feasible to achieve that 
  $y_d(t,\oo\eta_1,...,\oo\eta_d)= x(t_m+T)$ for all $m$, since it cannot be guaranteed  that   system (\ref{sys3})
  is solvable for   $\zeta_m\equiv  x(t_m+T)$: the system will be overdefined.  
  Nevertheless, estimate presented in  (\ref{eta1})  can still be achieved for any arbitrarily  large $\oo d$, since  (\ref{eta1}) holds.  A solution  could be found using  
methods for fitting linear models.

\comm{Furthermore, instead of calculation of the coefficients $a_0,a_1,...,a_d$, via solution of the approximation problem for the complex exponential described  in Theorem \ref{ThP}(i)-(ii), one may  find these coefficients considering them  as
additional unknowns  in system (\ref{sys3}) with $\oo d> 2d+1$.
Theorem \ref{ThP} implies again that there exist $\oo\eta_k=\eta_k\in\R$ and $a_k\in \R$ such that (\ref{sys3}) holds with $\zeta_m=\w y_d(t_m)$. This would lead to a 
nonlinear fitting problem for unknowns $a_0,a_1,...,a_d,\oo\eta_1,...,\oo\eta_d$.}

 So far,  the consistency of these procedures is unclear since  a choice of smaller $\e$ leads to larger $d$. We leave analysis of these methods for the future research. 
 \subsubsection*{The case of causal filtering problem: Theorem \ref{ThF} setting} 
In  the setting of Theorem \ref{ThF}, the past values of the true unknown process $\ww x(t)$ are not observable and hence  cannot be used for fitting the values $\eta_1,...,\eta_d$. However, we can use  that the values $\eta_1,...,\eta_k$ in (\ref{viaeta0})-(\ref{viaeta}) are still the same as in the setting of Theorem \ref{ThF}, where $\ww x(t)=x(t+T)$. Since past $x(s)|_{s=t_1,...,t}$ are observable,  we can use the fitting procedure  based on Theorem \ref{ThP}  to estimate  $\eta_1,...,\eta_d$ using (\ref{viaeta0})-(\ref{viaeta}) with the coefficients  $a_k$ defined for approximation of $\zeta(\o)=e^{i\o T}$ and with observations $x(t_m)$, $t_m\le t$,  as described above. After that, we can  estimate $\ww x(t)$ using  equation  (\ref{viaeta0}) again with the new coefficients  $a_k$ defined for approximation of $\zeta(\o)=\Ind_{|\o|\le \O}$.
\section{A possible choice of $\psi_d$ for predictors in Theorem \ref{ThP} setting}
\label{secExp}
The coefficients $a_k$ for functions $\psi_n$ could be found use numerical methods from classical analysis such as  the Gram-Schmidt procedure.  In the case of Theorem \ref{ThP} for predictors, finding these coefficients  can be simplified, especially for $T=1$.

 Let us demonstrate this.

Assume that $T=1$.  For real $\nu>0$, 
define on $\C\setminus\{1\}$ a function
\baaa
\ww\psi(\nu,z)\defi z\left(1-\exp\frac{\nu }{1-z}\right).
\eaaa
This function is a modification of the transfer  function introduced in \cite{D12} for prediction of signals with a single point spectrum degeneracy.
Clearly,
\baaa
\Re\frac{\nu }{1-e^{i\o}}=\nu\frac{1-\cos(\o)}{|1-e^{i\o}|^2}\to 0\quad\hbox{as}\quad \nu\to -\infty
\eaaa
uniformly on the set $\{e^{i\o}, \quad \o\in (-\pi,\pi],\ |\o|\ge\O\}$.
Hence \baaa
\psi(\nu,z)\to z\quad\hbox{as}\quad \nu\to -\infty
\eaaa
uniformly on the set $\{e^{i\o}, \quad \o\in (-\pi,\pi],\ |\o|\ge\O\}$. 

Further, for $\e>0$,  let $\nu<0$ be selected such that 
\baaa
|\psi(\nu,e^{i\o})-e^{i\o}|\le \frac{\e}{2},\quad   \o\in (-\pi,\pi],\ |\o|\ge\O.
\eaaa

The function $\ww\psi(\nu,\cdot)$ is analytic in $\C\setminus\{1\}$, and is bounded on $\C\setminus\{z\in\C:\ |1-z|>\d\}$ for any $\d>0$. Clearly, we have that
\baaa
\ww\psi(\nu,z)\defi z\left(1-\left(1+\frac{\nu }{1-z}+\frac{1}{2}\left[\frac{\nu }{1-z}\right]^2
+\frac{1}{3!}\left[\frac{\nu }{1-z}\right]^3+\cdots +\frac{1}{d!}\left[\frac{\nu }{1-z}\right]^d+\cdots
\right)\right)\\
=\lim_{d\to +\infty}\ww\psi_d(\nu,z),
\eaaa
where
\baaa
\ww\psi_d(\nu,z)\defi -\frac{\nu z}{1-z}-\frac{z}{2}\left[\frac{\nu }{1-z}\right]^2
-\frac{z}{3!}\left[\frac{-\nu }{1-z}\right]^3-\cdots -\frac{z}{d!}\left[\frac{-\nu }{1-z}\right]^d,
\eaaa
and where convergence is uniform on the set $\{e^{i\o}, \quad \o\in (-\pi,\pi],\ |\o|\ge\O\}$.

It can be observed that the functions $\psi_d(\nu,z)$ belong to $\Psi_d$, since 
\baaa
\frac{z}{1-z}=\frac{1}{1-z^{-1}},\quad \frac{1}{1-z}=1-\frac{1}{1-z^{-1}},\quad z\in\C,\quad z\neq 1.
\eaaa
 For example, 
 \baaa
\frac{z}{2}\left[\frac{\nu }{1-z}\right]^2=\frac{\nu^2}{2(1-z^{-1})}\left(1-\frac{1}{1-z^{-1}}\right).
\eaaa

 Clearly,  we can select  $d$ such that
 \baaa
|\psi(\nu,e^{i\o})-\psi_d(\nu,e^{i\o})|\le \frac{\e}{2},\quad   \o\in (-\pi,\pi],\ |\o|\ge\O.
\eaaa
 For this $d$ and $\nu$, we have that 
 \baaa
|\psi_d(\nu,e^{i\o})-e^{i\o}|\le \frac{\e}{2},\quad   \o\in (-\pi,\pi],\ |\o|\ge\O.
\eaaa
 The coefficients $a_k$ can be computed form the representation of $\psi_d$ as an element of $\Psi_d$.  
 
For the case of $T>1$, one can use functions $\psi_d(\nu,z)^T$.

\section{Low frequency  and other signals}\label{secLF}
Let  us show that the results obtained  above  for high frequency signals can be applied to signals of more general type described as follows. 

Let $\O\in  (0,\pi)$, $\O_0\in (0,\O)$, and $\t\in  (-\pi,\pi]$ be given, and let 
 $\Y(\O,\t)$   be the set of all signals 
$x\in\ell_2$ such that $X\left(e^{i(\o-\t)}\right)=0$ for $|\o|<\O$, $\o\in [-\pi,\pi]$ and  $X=\Z x$.

For example,  $\Y(\O,0)=\X(\O)$; this set includes high frequency signals  such that
$X\ew=0$ if $|\o|<\O$.  Respectively, the set $\Y(\O,\pi)$ includes low frequency signals (band limited signals) such that
$X\ew=0$ if  $\o\in (-\pi,-\pi+\O)\cup  (\pi-\O,\pi]$.

To predict a signal $\w x\in \Y(\O,\t)$, one can convert it into 
a signal $ x\in \X(\O)=\Y(\O,0)$ as $x(t)=e^{-i\t t}\w x(t)$. Then one can use for $x$ the predictors  introduced in Theorem \ref{ThP}.
 The implied prediction  $\w y(t)$ for $\w x(t)$ can be obtained as $\w y(t)=e^{i\t t}y(t)$,
 where  $y(t)$ is the corresponding  prediction  for $x(t)$.

Similarly, one can construct a causal filter that,  for $x\in \Y(\O_0,\t)$, produces
an approximation of  $\w x\in \Y(\O,\t)$ such that $\w x=\Z^{-1}\Phi_{\O,\t} X$, where $X=\Z x$, and $\Phi_{\O,\t}$ is Z-transform of an ideal filter such that   $\Phi_{\O,\t}\left(e^{i(\o-\t)}\right)=\Ind_{|\o|>\O,\ \o\in [-\pi,\pi]}$. 
Again, one can convert it into 
a signal $ x\in \X(\O_0)=\Y(\O_0,0)$ as $x(t)=e^{-i\t t}\w x(t)$. Then one can use for $x$ the causal filter  introduced in Theorem \ref{ThF}.
 The implied filtered signal $\w y(t)$ for $\w x(t)$ can be obtained as $\w y(t)=e^{i\t t}y(t)$,
 where  $y(t)$ is the corresponding  filtered  signal for $x(t)$.

Alternatively, we can construct predictors and filters  directly for  signals from $\Y(\O,\t)$
similarly to the ones introduced in Theorems \ref{ThP}-\ref{ThF} and with the transfer  functions 
\baaa
\sum_{k=0}^d \frac{a_k}{(1-e^{i\t}/z)^k}
\label{g1}\eaaa approximating $e^{i\o T}$ and $\Ind_{|\o|>\O}$ on $\TT$. 

In the setting where $x\in \cup_\t\Y(\O,\t)$,  and where $\t$ is unknown, we can use approach from Section \ref{secD} to fit $\t$ from past observations as a new unknown parameter. 

 \section{Proofs}\label{secProof}
{\em Proof of Lemma \ref{lemmaA}}.  
For a set  $I\subset [-\pi,0)\cup(0,\pi]$, 
let  $\g^E(I)$ (or $\g_d^E(I)$) be the set of functions $\g:I\to \C$ constructed as  $\g(\o)=\psi\ew$ for some $\psi$ from $\Psi^E$ (or from $\G_d^E(I)$, respectively)).  

Let  $I_{\oo\O}\defi [-\pi,-\oo\O]\cup  [\oo\O,\pi] $. 
\comm{Let function $\zeta:[-\pi,\p]\to \C$ be defined as $\zeta(\o)=e^{i\o T}$ for the proof of statement (i),
and   as $\zeta(\o)=\Ind_{|\o|\ge \O_0}$ for the proof of statement (ii).}

Clearly,  $\frac{1}{1-z^{-1}}=1-\frac{1}{1-z}$ for all $z\in\C$, $z\neq 1$.
Hence \baaa\overline{\left(\frac{1}{1-1/e^{i\o}}\right)}=\frac{1}{1-e^{i\o}}=1-\frac{1}{1-1/e^{i\o}}, \qquad \o\in\R,\quad  \o\neq 0.\eaaa
It follows that, if $\psi(z)\in\Psi^E$ then $\psi(z^{-1})\in\Psi^E$, for both $E=\R$ and $E=\C$. This implies that  $\overline{\g(\o)}\in \G^\E(I_{\oo\O})$ if $\g(\o)\in\G^E(I_{\oo\O})$. 

Since the function $\g_1(\o)=\Re\frac{1}{1-e^{-i\o}}$ is strictly monotone 
on the intervals $(-\infty,0)$ and  $(0,\infty)$, and has different signs on these two intervals,
it follows that $\g_1(\a)\neq \g_1(\b)$ for all  $\a,\b\in I_{\oo\O}$, $\a\neq \b$.  
It follows that
 the set of function
$\G^\C(I_\O)$ separates points on the compact set $I_{\oo\O}$.
By the Stone-Weierstrass Theorem for  complex valued continuous functions on compact sets of real numbers, it follows that the set $\G(I_{\oo\O})$ is complete in the space  $C(I_{\oo\O};\C)$ of continuous complex-valued functions defined on  $I_\O$ with the supremum norm;
see, e.g., Theorem 10 in \cite{SW}, pp. 238.  It follows that, for any $\ee>0$,  there exists $d>0$ and $\w\g_d\in \G_d^\C(I_{\oo\O})$  represented as 
 $\w\g_d(\o)=\sum_{k=0}^d \frac{A_k}{\left(1-e^{-i\o}\right)^{k}}$ defined for $\o\in\R\setminus \{0\}$, where $A_k\in \C$,  such that
\baaa
&& \left(\int_{I_{\oo\O}}|\zeta(\o)-\w\g_d(\o)|^2d\o\right)^{1/2}\le\ee.
\eaaa
For $\zeta(\o)=e^{i\o T}$ this follows directly from Theorem 10 in \cite{SW}, pp. 238, mentioned above. 
For $\zeta(\o)=\Ind_{|\o|\ge \oo\O}$ this follows from the fact that 
the set $C(I_\O;\C)$ is
everywhere dense in  $L_2(I_{\oo\O};\C)$, and convergence in   $C(I_{\oo\O};\C)$ implies  convergence in $L_2(I_{\oo\O};\C)$.   

Let us show that 
the same estimate holds for  $\g_d\in\G_d^\R(I_\O)$
defined as 
$\g_d(\o)=\sum_{k=0}^d \frac{a_k}{\left(1-e^{-i\o}\right)^{k}}$, where $a_k=\Re A_k$.

Suppose that  $\Im A_k\neq 0$ for some $k$. Clearly, 
the real and the the imaginary part of $\frac{ i \Im A_k}{\left(1-e^{-i\o}\right)^{k}}$ 
are even and odd, respectively. On the other hand, 
the functions $\Re e^{i\o T}=\cos(\o T)$ and  $\Im e^{i\o T}=\sin(\o T)$ are odd 
and even, respectively,  on $\R$. Therefore, the replacement of $A_k$ by $a_k=\Re A_k$
cannot spoil the estimate. 
Hence the   transfer 
function   $\psi_d\ew=\g_d(\o)=\sum_{k=0}^d \frac{a_k}{\left(1-e^{-i\o}\right)^{k}}$ satisfies the required estimate. This completes the prove of Lemma \ref{lemmaA}. $\Box$

    \vspace{2mm}
\par
{\em Proof of Theorems \ref{ThP}-\ref{ThF}}. 
Let us consider first the case where  $x\in \X(\infty,\O)=\X(\O)$.

We continue with the  proof   for  Theorems \ref{ThP} (with $\tau=+\infty)$ and Theorem \ref{ThF} simultaneously.   
For the proof of Theorem \ref{ThP},  we assume that  $\oo\O=\O$ and  $\zeta:I_\O\to \C$ is defined as $\zeta(\o)=e^{i\o T}$. For the proof of Theorem \ref{ThF}, we assume that
 $\oo\O=\O_0$  as $\zeta(\o)=\Ind_{|\o|\ge \O}$.

 Assume that estimate (\ref{e2}) holds for selected $d,\g_d,\psi_d$. We have that 
\baaa
&&\ww x(t)-\w y_d(t)=\int_{-\pi}^\pi e^{i \o t} (\zeta(\o)-\psi_d\ew)X\ew d\o.
\eaaa
where  $\ww x(t)=x(t+T)$  in the setting of Theorem \ref{ThP},  and $\ww x(t)$ is an ideal filtered process in the setting of Theorem \ref{ThF}. Clearly, 
 \baaa
|\ww x(t)-\w y_d(t)|&\le& \left(\int_{-\pi}^\pi |\zeta(\o)-\psi_d\ew|  d\o\right)^{1/2}
\\ &\le& \left(\int_{I_{\oo\O}}|\zeta(\o)-\psi_d\ew|^2 d\o\right)^{1/2}
\left(\int_{-\pi}^\pi|X\ew|^2 d\o\right)^{1/2} \le \ee.
\eaaa
We have that
\baaa
\left(\int_{-\pi}^\pi|X\ew|^2 d\o\right)^{1/2}=\|x\|_{\ell_2}.
\eaaa
Hence $|\ww x(t)-\w y_d(t)|\le \ee \|x\|_{\ell_2}$. This implies  the proofs of Theorems \ref{ThP} for the case where $\tau=+\infty$ and  Theorem \ref{ThF}. 

Let us prove Theorem \ref{ThP} for the case where $\tau=0$.
Let $x\in \X(0,\O)$.
 Let us define an even function $\ww x:\ZZ\to \R$ such that $\ww x(t)=x(t)$ for $t\le 0$, and
 $\ww x(t)=x(-t)$ for $t>0$. Let   $\ww X=\Z \ww x$. 
 It can be shown that $\Re \ww X\ew=2\sum_{t=-\infty}^{-1} \cos(\o t)x(t)+x(0)$ and $\Im \ww X\ew=0$ for $\o\in(-\pi,\pi]$.
This implies that $\ww x\in \XHF$. Furthermore, since predictors $p_d$ are causal, it follows that $p_d(\ww x(\cdot))(t)=p_d(x(\cdot))(t)$ for all $t\le 0$. 
Then the proof for $\tau=0$ follows from the proof  the case of $\tau=+\infty$. 

The case where $\tau=-1$ can be considered similarly. For $x\in \X(-1,\O)$, we define an odd function $\ww x:\ZZ\to\R$ such that $\ww x(t)=x(t)$ for $t\le -1$, $\ww x(0)=0$, and 
 $\ww x(t)=-x(-t)$ for $t>0$. Let  $\ww X=\Z \ww x$. 
 It can be shown that $\Re \ww X\ew=0 $ and $\Im \ww X\ew=2\sum_{t=-\infty}^{-1} \sin(\o t)\ww x(t)$ for $\o\in(-\pi,\pi]$. It follows that $\ww x\in \XHF$. Again, since predictors the $p_d$ are causal, it follows that $p_d(\ww x(\cdot))(t)=p_d(x(\cdot))(t)$ for all $t\le -1$.
Hence the  proof for $\tau=-1$ follows from the proof for the case of $\tau=+\infty$.
 This completes the proofs of Theorems \ref{ThP}. 
$\Box$

{\em Proof of Lemma \ref{lemma1}}.
We have that
\baaa x_k(t)=\eta_k+\sum_{s=t_1}^{t}x_{k-1}(s)=
\sum_{l=1}^k c_{l}(t)\eta_l+f_{k}(t).
\label{eta22}\eaaa
Futrher, we have  that   
 $y_d(t)=a_0x(t)+\sum_{k=1}^d a_k x_k(t)$ for any $t\ge t_1$, i.e.,
  \baa
y_d(t) =a_0x(t)+\sum_{k=1}^d a_k \left(\eta_k+\sum_{s=t_1}^{t}x_{k-1}(s)\right).
 \label{sys} \eaa
Here we assume that $x_0\defi x$.

Furthermore, we have  that  
 \baaa
 &&\sum_{\tau=t_1}^{t_{}}x_{1}(\tau) =\sum_{\tau=t_1}^{t_{}}\left(\eta_1 +\sum_{s=t_1}^\tau x_{0}(s)\right)\brea=\eta_1(t_{}-t_1+1)+\sum_{\tau=t_1}^{t_{}}\sum_{s=t_1}^\tau x_{0}(s)
\comm{ \\&&=\eta_1(t_{}-t_1+1) +\sum_{s=t_1}^{t_{}}\sum_{\tau=s}^t x_{0}(s) 
 =\eta_1(t-t_1+1)+\sum_{s=t_1}^t x_{0}(s)(t-s+1)}
 \eaaa
 and
 \baaa 
\sum_{\tau_1=t_1}^{t_{}}x_{2}(\tau_1) =\sum_{\tau_1=t_1}^{t_{}}\left(\eta_2 +\sum_{s=t_1}^{\tau_1} x_{1}(s)\right)\brea=\eta_2(t_{}-t_1+1)+
\sum_{\tau_1=t_1}^{t_{}}\sum_{s=t_1}^{\tau_1} x_{1}(s)\\
  =
 \eta_2(t_{}-t_1+1)+\sum_{\tau_1= t_1}^{t_{}}  
\left[\eta_1(\tau_1-t_1+1)+  \sum_{\tau_2=t_1}^{\tau_1}\sum_{s= t_1}^{\tau_2} x_{0}(s)\right]
\\= 
 \eta_2(t_{}-t_1+1)+\eta_1 \sum_{\tau_1= t_1}^{t_{}}   (\tau_1-t_1+1)
+\sum_{\tau_1=t_1}^{t_{}}      \sum_{\tau_2=t_1}^{\tau_1}\sum_{s= t_1}^{\tau_2} x_{0}(s).
 \eaaa
 Similarly, 
 \baaa 
\sum_{\tau_1=t_1}^{t_{}}x_{3}(\tau_1) =\sum_{\tau_1=t_1}^{t_{}}\left(\eta_3 +\sum_{s=t_1}^{\tau_1} x_{2}(s)\right)\brea=\eta_3(t_{}-t_1+1)+
\sum_{\tau_1=t_1}^{t_{}}\sum_{s=t_1}^{\tau_1} x_{2}(s)
\\
  = 
\eta_3(t_{}-t_1+1)+\sum_{\tau_1= t_1}^{t_{}}  
\left[\eta_2(\tau_1-t_1+1)+  \sum_{\tau_2=t_1}^{\tau_1}\sum_{s= t_1}^{\tau_2} x_{1}(s)\right]
\\= 
 \eta_3(t_{}-t_1+1)+\eta_2 \sum_{\tau_1= t_1}^{t_{}}   (\tau_1-t_1+1)
+\sum_{\tau_1=t_1}^{t_{}}  \sum_{\tau_2=t_1}^{\tau_1} \sum_{\tau_3=t_1}^{\tau_2} x_{1}(\tau_3)\\
= \eta_3(t_{}-t_1+1)+\eta_2 \sum_{\tau_1= t_1}^{t_{}}  (\tau_1-t_1+1)
+
\sum_{\tau_1=t_1}^{t_{}}  \sum_{\tau_2=t_1}^{\tau_1} \sum_{\tau_3=t_1}^{\tau_2}\Bigl[\eta_1+\sum_{s=t_1}^{\tau_3}x_{0}(s)\Bigr]
\\= 
\eta_3(t_{}-t_1+1)+\eta_2 \sum_{\tau_1= t_1}^{t_{}}  (\tau_1-t_1+1)
+
\sum_{\tau_1=t_1}^{t_{}}   \sum_{\tau_2=t_1}^{\tau_1}\Bigl[\eta_1 (\tau_2-t_1+1)
+\sum_{s=t_1}^{\tau_2}\sum_{s=t_1}^{\tau_3}x_{0}(s)\Bigr]
\\=
\eta_3(t_{}-t_1+1)+\eta_2 \sum_{\tau_1= t_1}^{t_{}}  (\tau_1-t_1+1)
+
\sum_{\tau_1=t_1}^{t_{}}   \sum_{\tau_2=t_1}^{\tau_1}\eta_1 (\tau_2-t_1+1)
+
\sum_{\tau_1=t_1}^{t_{}}   \sum_{\tau_2=t_1}^{\tau_1}\sum_{s=t_1}^{\tau_2}\sum_{s=t_1}^{\tau_3}x_{0}(s)
 \eaaa
 
 Similarly, we obtain that, for $k>2$,
\baaa
 \sum_{s=t_1}^{t_{}}x_{k}(s) = \eta_k(t_{}-t_1+1)+\eta_{k-1} 
\sum_{\tau_1= t_1}^{t_{}}   (\tau_1-t_1+1)+...\breakk +
 \eta_{1} \sum_{\tau_1=t_1}^{t_{}} \sum_{\tau_2=t_1}^{\tau_1} ...\sum_{\tau_k=t_1}^{\tau_{k-1}}(\tau_{k}-t_1+1)
 \\
+\sum_{\tau_1=t_1}^{t_{}} \sum_{\tau_2=t_1}^{\tau_1} ...\sum_{s=t_1}^{\tau_k} x_{0}(s).
 \label{etak}\eaaa
\par
It follows that
\baaa x_k(t)=\eta_k+\sum_{s=t_1}^{t}x_{k-1}(s)=
\sum_{l=1}^k c_{l}(t)\eta_l+f_{k}(t).
\label{eta222}\eaaa
Together with (\ref{sys}), this proves (\ref{viaeta}) and completes the proof of Lemma \ref{lemma1}. $\Box$
\section{Concluding remarks}
\begin{enumerate}
\item
The  approach suggested in this paper  allows many modifications. 
In particular,   other non-causal discrete time transfer functions can be approximated by 
causal transfer functions from $\Psi^E$.  In fact, any transfer function $H(z)$ can be approximated that way if $\int_{-\pi}^\pi |H\ew|^2d\o <+\infty$. 

\item
 It can be shown that, by  Theorem 10 in \cite{SW}, pp. 238 again, 
approximation of $\zeta(\o)=\Ind_{|\o|\ge
  \oo\O}$ in Lemma \ref{lemmaA}  can 
  be in fact achieved on the set of real valued functions represented as
 \baaa
 \g_d(\o)=\psi_d\ew=\sum_{k=0}^db_k\left|\frac{1}{1-e^{i\o}}\right|^{2k}=\sum_{k=0}^db_k\left(\frac{1}{1-e^{-i\o}}\right)^k \left(1-\frac{1}{1-e^{-i\o}}\right)^k
 \eaaa
 with $b_k\in\R$. This may help to streamline calculations since this set  is smaller  than $\Psi^R$. If $b_k$ are found, then we can derive the  coefficients $a_k$ needed  for the  fitting of $\eta_k$ via (\ref{viaeta0})-(\ref{viaeta}).
 \item The predictors introduced in \cite{D12,D12b} do not allow the fitting procedure described in Section \ref{secD} since the kernels of the corresponding causal convolutions are heavily time dependent. 
 \item In the present paper, we consider $L_2$-approximation  of non-causal transfer functions; this allowed to approximate discontinuous on $\TT$ transfer functions  used for the filtering problem. In addition, this would allow to use  the Gram-Schmidt  procedure to construct  the functions $\psi_d$. 
 This was not feasible in the continuous time setting \cite{D22rus}, where the uniform approximation on the infinite intervals  was required. 
 \item In general, it can be expected that the approximating functions $\psi_d\ew$ take large
 values for large $d$ inside the interval $(-\oo\O,\oo\O)$, in the terms of Lemma \ref{lemmaA}. However,  some robustness of the prediction and filtering with respect to noise contamination can be established similarly to \cite{D12}. We leave it for the future research.
 \item The processes from $\X(\tau,\O)$  do not necessarily  have a spectrum degeneracy for $\tau=-1,0$ and $\O\in (0,\pi)$; in fact,  their Z-transforms can be separated from zero on $\TT$. However,
Theorem \ref{ThP} shows that they are predictable on the left half of the timeline because of  their  left-sided  spectrum degeneracy defined by (\ref{L1}),(\ref{L2}). 
\end{enumerate}


\begin{thebibliography}{100}
\bibitem{B}
Butzer, P.L., Stens, R.L. (1993). Linear Prediction by Samples from the Past. In: Marks, R.J. (eds) {\em Advanced Topics in Shannon Sampling and Interpolation Theory. \comm{Springer Texts in Electrical Engineering.}} Springer, New York, NY.
\bibitem{Higgins96}
Higgins, J.R. (1996). {\em Sampling Theory in Fourier and Signal
Analysis}. Oxford University Press, New York.
\bibitem{LH}
Li, Z., Han, J., Song, Yu. J. (2020).
On the forecasting of high-frequency financial time series based on ARIMA model
improved by deep learning. {\em J. of Forecasting } 39(7), 1081--1097
\bibitem{LT}
Luo, S.,  Tian, C,  (2020). Financial high-frequency time series forecasting based on sub-step grid search long short-term memory network, {\em IEEE Access}, Vol. 8, 203183 - 203189.
\bibitem{K}
Knab J.J. (1979). Interpolation of band-limited functions using
the approximate prolate series. {\it IEEE Transactions on
Information Theory} {\bf 25}(6), 717--720.
\bibitem{L}
Lyman R.J, Edmonson, W.W., McCullough S., and Rao M. (2000). The
predictability of continuous-time, bandlimited processes. {\it
IEEE Transactions on Signal Processing} {\bf 48}(2), 311--316.
\bibitem{Ly}
Lyman R.J and Edmonson, W.W. (2001). Linear prediction of
bandlimited processes with flat spectral densities. {\it IEEE
Transactions on Signal Processing} {\bf 49} (7), 1564--1569.
\comm{\bibitem{M}
Marvasti F. (1986). Comments on "A note on the predictability of
band-limited processes." {\it Proceedings of the IEEE} {\bf
74}(11), 1596.}
\bibitem{P}
Papoulis A. (1985). A note on the predictability of band-limited
processes. {\it Proceedings of the IEEE} {\bf 73}(8), 1332--1333.
\bibitem{V}
 Vaidyanathan P.P. (1987). On predicting a band-limited signal
based on past sample values. {\it Proceedings of the IEEE} {\bf
75}(8), 1125--1127.
\bibitem{D12}
 Dokuchaev, N. (2012). Predictors for discrete time processes with
energy decay on higher frequencies. {\em IEEE
Transactions on Signal Processing} {\bf 60}, No. 11, 6027-6030.
\bibitem{D12b} Dokuchaev, N. (2012). On predictors for band-limited and
high-frequency time series. {\em Signal Processing} {\bf 92}, iss.
10, 2571-2575.
\bibitem{D16} Dokuchaev, N. (2016). Near-ideal causal smoothing filters for the real sequences. {\em Signal Processing} {\bf 118}, iss. 1, pp.285-293.
\bibitem{D22}
Dokuchaev, N. (2022). Limited memory predictors based on polynomial approximation of periodic
exponentials. {\em Journal of Forecasting}
41 (5), 1037-1045.
\bibitem{D22rus}
Dokuchaev, N. (2022). Predictors for high frequency signals based on rational polynomials approximation of periodic exponentials. {\em Problems of Information Transmission} 58, No. 4, pp. 372–381.
\bibitem{SW}
Stone, M.H. (1948). The generalized Weierstrass approximation theorem. {\em Mathematics  Magazine}
Vol. 21, \comm{No. 4, pp. 167-184,} No. 5, 237-254.
\end{thebibliography}
\end{document}